\documentclass[english,aps,preprintnumbers,nofootinbib,superscriptaddress]{revtex4-1}
\usepackage{amssymb,amsmath}
\usepackage{graphicx}

\begin{document}
\preprint{YITP-18-55, IPMU18-0092}
\title{Vector disformal transformation of generalized Proca theory}

\author{Guillem Dom\`enech}
\email[]{g.domenech-AT-thphys.uni-heidelberg.de}
\affiliation{Institute for Theoretical Physics, Heidelberg University, Philosophenweg 16
D-69120 Heidelberg, Germany}

\author{Shinji Mukohyama}
\email[]{shinji.mukohyama-AT-yukawa.kyoto-u.ac.jp}
\affiliation{Center for Gravitational Physics, Yukawa Institute for Theoretical Physics, Kyoto University, 606-8502, Kyoto, Japan}
\affiliation{Kavli  Institute  for  the  Physics  and  Mathematics  of  the  Universe  (WPI), The  University  of  Tokyo  Institutes  for  Advanced  Study, The  University  of  Tokyo,  Kashiwa,  Chiba  277-8583,  Japan}

\author{Ryo Namba}
\email[]{namba-AT-physics.mcgill.ca}
\affiliation{Department of Physics, McGill University, Montr\'{e}al, QC, H3A 2T8, Canada}

\author{Vassilis Papadopoulos}
\email[]{papadopo-AT-clipper.ens.fr}
\affiliation{Ecole Normale Superieure, 45 rue d'Ulm,75005 Paris, France}

\date{\today}

\begin{abstract}
 Motivated by the GW170817/GRB170817A constraint on the deviation of the speed of gravitational waves from that of photons, we study disformal transformations of the metric in the context of the generalized Proca theory. The constraint restricts the form of the gravity Lagrangian, the way the electromagnetism couples to the gravity sector on cosmological backgrounds, or in general a combination of both. Since different ways of coupling matter to gravity are typically related to each other by disformal transformations, it is important to understand how the structure of the generalized Proca Lagrangian changes under disformal transformations. For disformal transformations with constant coefficients we provide the complete transformation rule of the Lagrangian. We find that additional terms, which were considered as beyond generalized Proca in the literature, are generated by the transformations. Once these additional terms are included, on the other hand, the structure of the gravity Lagrangian is preserved under the transformations. We then derive the transformation rules for the sound speeds of the scalar, vector and tensor perturbations on a homogeneous and isotropic background. We explicitly show that they transform following the natural expectation of metric transformations, that is, according to the transformation of the background lightcone structure. We end by arguing that inhomogeneities due to structures in the universe, e.g.~dark matter halos, generically changes the speed of gravitational waves from its cosmological value. Therefore, even if the propagation speed of gravitational waves in a homogenoues and isotropic background is fine-tuned to that of light (at linear level), the model is subject to further constraints (at non-linear level) due to the presence of inhomogeneities. We give a rough estimate of the effect of inhomogeneities and find that the fine-tuning should not depend on the background or that the fine-tuned theory has to be further fine-tuned to pass the tight constraint. 
\end{abstract}

\maketitle

\section{Introduction}

The recent multi-messenger detection of a binary neutron star merger, GW170817~\cite{TheLIGOScientific:2017qsa} and GRB170817A~\cite{Monitor:2017mdv}, put a stringent constraint on the deviation of the propagation speed of gravitational waves $c_{\rm T}$ from that of photons $c_{\gamma}$, as
\begin{equation}
 -3\times 10^{-15}\lesssim \delta c_{\rm T} \lesssim 7\times 10^{-16}\,, \label{eqn:GW170817constraint}
\end{equation}
where $\delta c_{\rm T}\equiv (c_{\rm T}-c_{\gamma})/c_{\gamma}$. This has significant implications to some modified gravity theories in which $\delta c_{\rm T}$ can be non-zero, depending on the choice of parameters and backgrounds~\cite{Creminelli:2017sry,Ezquiaga:2017ekz,Gumrukcuoglu:2017ijh,Crisostomi:2017pjs,Baker:2017hug,Sakstein:2017xjx,Langlois:2017dyl,Oost:2018tcv,Amendola:2018ltt,deRham:2018red}. From the phenomenological viewpoint, the new constraint narrows down the observationally viable range of parameters from an otherwise vast parameter space. From the theoretical viewpoint, it motivates a search for a consistent framework that renders a small $\delta c_{\rm T}$ (technically) natural. From either point of view, it is important to understand implications of the constraint \eqref{eqn:GW170817constraint} itself and the nature of $\delta c_T$ in a given theory confronted with observations.

The quantity $\delta c_{\rm T}$ depends on both the Lagrangian of the gravity sector and the way the matter sector (including the electromagnetism) is coupled to the gravity sector. One can therefore consider the constraint as a restriction on the gravity Lagrangian, the matter coupling to gravity or a combination of them. For this reason it is important to see how the gravity Lagrangian changes under metric transformations that relate different matter couplings, and how $c_{\rm T}$ changes accordingly. In the present paper we investigate these questions in the context of the generalized Proca theory. 

The generalized Proca theory~{\cite{Tasinato:2014eka,Heisenberg:2014rta,Allys:2015sht,Jimenez:2016isa,Allys:2016jaq}} is a vector-tensor theory of gravity that propagates five local degrees of freedom, two from a massless graviton and three from a massive vector field. Unlike a scalar field, a vector field can easily mediate a repulsive force and thus might weaken gravity in late-time cosmology to help reduce some reported tension between early-time and late-time cosmology~\cite{DeFelice:2016uil} (see \cite{DeFelice:2016ufg} for the possibility of weaker late-time gravity in the context of a massive gravity theory). Among gravity theories with a vector field, the generalized Proca theory is one of the interesting models that has second-order equations of motion and thus avoids ghosts associated with higher-derivative terms, also referred to as the Ostrogradsky ghosts \cite{Ostrogradsky:1850fid}~\footnote{In the generalized Proca theory, variation of its action immediately leads to second-order differential equations. However, in the cases where the kinetic matrix is degenerate, a more general action can be constructed by eliminating the Ostrogradsky ghosts thanks to additional constraints. Scalar-tensor theories of this type are called Degenerate Higher Order Scalar Tensor (DHOST) theories \cite{Langlois:2015cwa,BenAchour:2016fzp}, and a similar consideration for vector-tensor theories has been done in \cite{Kimura:2016rzw}.}.
However, the Lagrangian of the theory includes many free functions. It is important to see how the constraint (\ref{eqn:GW170817constraint}) restricts those free functions and/or the way {how} matter sector couples to the gravity sector. We shall therefore investigate the {behavior} of the generalized Proca Lagrangian under a certain class of metric transformations, to identify the degeneracy between {implications} of the constraint (\ref{eqn:GW170817constraint}) to the gravity Lagrangian and {those} to the matter coupling. 

For concreteness, in the present paper we restrict our consideration to a class of metric transformations called disformal transformations~\cite{Bekenstein:1992pj}. 
{This transformation was first introduced for a scalar field \cite{Bekenstein:1992pj} and has been extended to cases of a vector field \cite{Jimenez:2016isa,Kimura:2016rzw,Papadopoulos:2017xxx}.} Furthermore, for simplicity\footnote{{Otherwise we would end up with beyond generalized Proca theories.}} we only consider constant coefficients for the disformal transformations, namely of the form
\begin{equation}
 g_{\mu\nu} \rightarrow \mbox{(const.)}\times g_{\mu\nu} + \mbox{(const.)}\times A_{\mu}A_{\nu}\,, \label{eqn:constantdisformaltr}
\end{equation}
 where $g_{\mu\nu}$ is the metric and $A_{\mu}$ is the generalized Proca field. See \cite{Papadopoulos:2017xxx} for a similar disformal transformation by a gauge-invariant vector field. With this simplification, in the present paper we shall show the following two main statements.
\begin{itemize}
 \item[(a)] The constant disformal transformation (\ref{eqn:constantdisformaltr}) maps a generalized Proca Lagrangian to another generalized Proca Lagrangian. In showing this, we find a new term that should be included in the generalized Proca Lagrangian but that to our knowledge was considered to belong {only to} a wider class of theories \cite{Heisenberg:2016eld}.
 \item[(b)] The propagation speed of gravitational waves, which is written in terms of functions in the generalized Proca Lagrangian and {a} background configuration, changes under {a} constant disformal transformation (\ref{eqn:constantdisformaltr}) in a way that is easily inferred from the change of the background lightcone structure. We shall then extend this result to the vector perturbation and the scalar perturbation. 
\end{itemize}
Furthermore, in the present paper we shall make the following argument.
 \begin{itemize}
  \item[(c)]  Inhomogeneities due to structures in the universe such as galaxies and clusters induce additional contributions to $\delta c_{\rm T}$ and these environmental effects can be rather large. It is therefore not sufficient to satisfy the constraint (\ref{eqn:GW170817constraint}) for linearized gravitational waves around a homogeneous, isotropic cosmological background by fine-tuning. Generically, further fine-tuning {in relation to couplings to matter sector} is required, unless the cancellation is independent of the background \cite{Langlois:2017dyl}.
\end{itemize}

The rest of the present paper is organized as follows. In section~\ref{sec:Proca}, after briefly describing the generalized Proca theory, we shall show the statement (a) above and present the explicit transformation rules of the generalized Proca Lagrangian under constant disformal {transformations \eqref{eqn:constantdisformaltr}}. In section~\ref{sec:soundspeeds} we show the statement (b) above by computing the speed of gravitational waves (as well as those of vector and scalar perturbations) around a homogeneous, isotropic cosmological background before and after the transformations. In section~\ref{sec:constraint} we make some crude estimates of extra contributions to $\delta c_{\rm T}$ from inhomogeneities such as galaxies and clusters, and then make the argument (c) above. Finally, section~\ref{sec:summary} is devoted to a summary of the paper and discussions.

\section{Generalized Proca action and metric transformation}
\label{sec:Proca}

In this section, we first introduce the action of the generalized Proca theory \cite{Tasinato:2014eka,Heisenberg:2014rta,Allys:2015sht,Jimenez:2016isa,Allys:2016jaq}, a theory of one vector field without gauge invariance and non-minimally coupled to gravity. Although this theory contains derivative terms of the vector field in the action, it is constructed in such a way that the variation of the action does not generate pathological modes associated with higher-derivative terms in the equations of motion, usually referred to as Ostrogradsky ghosts \cite{Ostrogradsky:1850fid}. We are particularly interested in how sound speeds of perturbations on cosmological backgrounds transform under a disformal transformation of the metric. In this section, we derive the transformation rules of the generalized Proca action without specifying any background geometry. We focus on constant conformal and disformal factors as in \eqref{eqn:constantdisformaltr} and explicitly show that the generalized Proca action is closed under this transformation. We summarize our procedures and results for the transformation of the theory in the subsequent subsections, followed by Section \ref{sec:soundspeeds}, where we consider the transformation of sound speeds on cosmological backgrounds specifically.

\subsection{Model description}
\label{subsec:model}

We study a modified gravitational action obtained by allowing coupling of the metric $g_{\mu\nu}$ to a vector field $A_\mu$. In general, the Lagrangian of such a theory can be written in terms of $g_{\mu\nu}$, $A_\mu$, the covariant derivative $\nabla_\mu$ compatible with $g_{\mu\nu}$, the Riemann tensor $R_{\mu\nu\rho\sigma}$, the Levi-Civita tensor $\epsilon^{\mu\nu\rho\sigma}$, and their possible contractions. In order to avoid the presence of unwanted pathological modes that make the Hamiltonian of the system unbounded from below, the so-called Ostrogradsky instabilities \cite{Ostrogradsky:1850fid}, one can restrict the form of the action so that its variation gives rise to equations of motion only up to second order in derivatives acted on $g_{\mu\nu}$ and $A_{\mu}$. We consider an action of this type, called the generalized Proca action, which reads \cite{Heisenberg:2014rta,Allys:2015sht,Jimenez:2016isa,Allys:2016jaq} :
\begin{align}
S_{\rm GP} &= \sum_{i=2}^6 S_i \; , \quad 
S_i = \int d^4x \sqrt{-g} \, \mathcal{L}_i \; ,
\label{eq:genProc}\\
\mathcal{L}_2 &= G_2(X,F,U,Y) \; , 
\label{ProcL2}\\
\mathcal{L}_3 &= G_3(X)\nabla_{\mu}A^{\mu} \; ,
\label{ProcL3} \\
\mathcal{L}_4 &= G_4(X)R+G_{4,X}(X) \left[ \left( \nabla_{\mu} A^{\mu} \right)^2 - \nabla_{\rho} A^{\sigma} \nabla_{\sigma}A^{\rho} \right] \; ,
\label{ProcL4} \\
\mathcal{L}_5 &= G_5(X) \, G_{\mu\nu}\nabla^\mu A^\nu-\frac{1}{6} \, G_{5,X}(X) \left[ \left( \nabla_\mu A^\mu \right)^3 - 3\nabla_\rho A^\sigma \nabla_\sigma A^\rho \nabla_\mu A^\mu + 2 \nabla_\mu A^\nu \nabla_\rho A^\mu \nabla_\nu A^\rho \right] 
\nonumber \\
& \quad {-} g_5(X) \tilde{F}^{\mu\rho} \tilde{F}_{\nu\rho} \nabla_\mu A^\nu
{-} \mathcal{G}_5(X) \tilde{F}^{\mu\rho} \tilde{F}_{\nu\sigma} A_\mu A^\nu \nabla_\rho A^\sigma
 \; , 
\label{ProcL5} \\
\mathcal{L}_6 &= G_6(X) L^{\mu\nu\rho\sigma} \nabla_{\mu}A_\nu \nabla_{\rho}A_\sigma 
+ \frac{1}{2} \, G_{6,X}(X) \tilde{F}^{\mu\rho} \tilde{F}_{\nu\sigma} \nabla_\mu A^\nu \nabla_\rho A^\sigma \; ,
\label{ProcL6}
\end{align}
where $F_{\mu\nu} \equiv \nabla_\mu A_\nu - \nabla_\nu A_\mu$ and $\tilde{F}^{\mu\nu} \equiv \epsilon^{\mu\nu\rho\sigma} F_{\rho\sigma} / 2$ are respectively the field-strength tensor of the vector field and its dual, the subscript ``$, X$'' denotes derivative with respect to $X$, and various quantities appearing above are defined as
\begin{equation}
X \equiv - \frac{1}{2} \, A_\mu A^\mu \; , \quad
F \equiv - \frac{1}{4} F_{\mu\nu} F^{\mu\nu} \; , \quad
U \equiv - \frac{1}{4} F_{\mu\nu} \tilde F^{\mu\nu} \; , \quad
Y \equiv F_{\mu\rho} F_{\nu} {}^\rho A^\mu A^\nu \; , \quad
L^{\mu\nu\rho\sigma} \equiv \frac{1}{4} \, \epsilon^{\mu\nu\alpha\beta} \epsilon^{\rho\sigma\gamma\delta} R_{\alpha\beta\gamma\delta} \; .
\label{def-XFUY}
\end{equation}
In order to preserve parity, $G_2$ must be even in $U$, i.e. $G_2(X,F,U,Y)=G_2(X,F,-U,Y)$. Similarly to the action of the Horndeski theory, different terms in each ${\cal L}_i$ are tuned with each other so that the equations of motion remain second-order. Indeed, if we substitute $A_\mu = \nabla_\mu\phi$ into Eq.~\eqref{eq:genProc}, we recover the Horndeski action. The main difference comes from the purely antisymmetric terms such as $F_{\mu\nu}$, the second line in $\mathcal{L}_5$ and $\mathcal{L}_6$. Such terms vanish for $A_\mu = \nabla_\mu\phi$. In $\mathcal{L}_5$, we have included the term $\mathcal{G}_5(X) \tilde{F}^{\mu\rho} \tilde{F}_{\nu\sigma} A_\mu A^\nu \nabla_\rho A^\sigma$, which is usually omitted from the generalized Proca theory, only to be included in the ``beyond generalized Proca theory'' \cite{Heisenberg:2016eld}. However, one can verify that this term produces second-order equations of motion, and hence it could be included already. Moreover, as we shall see in Sec.~\ref{subsec:L5}, the disformal transformation of the $G_5$ and $G_{5,X}$ terms gives rise to terms of this form, and thus they are necessary in order to close the system under the transformation.

The goal of this section is to find the transformation rules due to the disformal transformation of the metric that is of type
\begin{equation}
 \bar{g}_{\mu\nu} = \Omega^2 \left[ \, g_{\mu\nu} + B A_\mu A_\nu \right]\,,\quad
  \Omega  = {\rm const.}\,,\quad B = {\rm const.}\,,
\label{disformal}
\end{equation}
where, from here on, we distinguish quantities in the two frames by the presence or absence of the upper ``bar.'' The conformal $\Omega$ and disformal $B$ factors could in principle depend on $X$, which is the only scalar quantity that can be constructed from $A_\mu$ and $g_{\mu\nu}$ without derivatives. However, in the case of $\partial \Omega / \partial X \ne 0$ or $\partial B / \partial X \ne 0$, the structure of the action \eqref{eq:genProc} without further generalization is not closed under the transformation \eqref{disformal} \cite{Kimura:2016rzw}. For simplicity we thus restrict our consideration to constant factors, and in the following subsections we explicitly prove that the action \eqref{eq:genProc} is indeed closed under the transformation \eqref{disformal} with this restriction. We assume that $A_\mu$ is independent of the transformation and stays as the same quantity in both frames. In other words, we consider $A_{\mu}dx^{\mu}$ (instead of $A^{\mu}\partial_{\mu}$) as fundamental.

The subsequent procedure of investigation is straightforward. We start with the action \eqref{eq:genProc} in the ``barred'' frame, apply the replacement \eqref{disformal}, and re-express the action in terms of ``unbarred'' quantities. Schematically, it reads
\begin{equation}
\bar{S}_{\rm GP} \left[ \bar{g}_{\mu\nu} \, , \, \bar{A}_\mu \right]
= \bar{S}_{\rm GP} \left[ \Omega^2 \left( g_{\mu\nu} + B A_\mu A_\nu \right) , \, A_\mu \right] \equiv S_{\rm GP} \left[ g_{\mu\nu} , A_{\mu} \right] \; .
\end{equation}
As we show below, it turns out that, by starting from $\bar{S}_{\rm GP}$ of the form \eqref{eq:genProc}, $S_{\rm GP}$ can also be written in the same form, only with redefinitions of the functions $G_{2,3,4,5,6}$, $g_5$ and ${\cal G}_5$. This is what we mean by ``an action closed under a transformation''. Let us repeat that the closure of $\bar{S}_5$ holds only if the ${\cal G}_5(X)$ term is present in \eqref{ProcL5}.

We now proceed to the computation of how each of $S_i$ in \eqref{eq:genProc} transforms under \eqref{disformal} with constant coefficients. To prepare, it is useful to know a few transformation rules of elementary quantities. The inverse and determinant of the metric transform as
\begin{equation}
\bar g^{\mu\nu} = \frac{1}{\Omega^2} \left( g^{\mu\nu} - \frac{B}{1 - 2 B X} A^\mu A^\nu \right) \; , \quad
\sqrt{- \bar{g}} = \Omega^2 \sqrt{ 1 - 2 B X} \sqrt{-g} \; ,
\label{ginv_det}
\end{equation}
the field-strength tensor and its dual as
\begin{equation}
\bar F_{\mu\nu} = F_{\mu\nu} \; , \quad
\bar{\tilde F}^{\mu\nu} = \frac{\tilde F^{\mu\nu}}{\Omega^4 \sqrt{1 - 2 B X}} \; , 
\label{F_Fdual}
\end{equation}
and the various scalar quantities as
\begin{equation}
\bar X = \frac{X}{\Omega^2 \left( 1 - 2 B X \right)} \; ,\quad
\bar F = \frac{1}{\Omega^4} \left[ F + \frac{BY}{2 (1 - 2 B X)} \right] \; , \quad
\bar U = \frac{U}{\Omega^4 \sqrt{1 - 2 B X}} \; , \quad
\bar Y = \frac{Y}{\Omega^6 \left( 1 - 2 B X \right)^2} \; .
\label{XFUY}
\end{equation}
The right-hand sides of the above equations are constructed solely by the ``unbarred'' metric $g_{\mu\nu}$. From the first equation in \eqref{XFUY}, it is immediate to see
\begin{equation}
\frac{\partial \bar{X}}{\partial X} = \frac{1}{\Omega^2 \left( 1 - 2 B X \right)^2} \; .
\label{dXbarX}
\end{equation}
Also, the derivative of $\bar A_\mu$ and the Riemann tensor transform as
\begin{equation}
\bar{\nabla}_\mu \bar{A}_\nu = \nabla_\mu A_\nu - \delta\Gamma^\rho_{\mu\nu} A_\rho \; , \qquad
\bar{R}^\rho{}_{\sigma\mu\nu} = R^\rho{}_{\sigma\mu\nu} 
+ \nabla_\mu \delta\Gamma^\rho_{\nu\sigma} - \nabla_\nu \delta\Gamma^\rho_{\mu\sigma} 
+ \delta\Gamma^\rho_{\mu\lambda} \delta\Gamma^\lambda_{\nu\sigma}
- \delta\Gamma^\rho_{\nu\lambda} \delta\Gamma^\lambda_{\mu\sigma} \; , 
\label{DA_Riemann}
\end{equation}
where
\begin{equation}
\delta \Gamma_{\mu\nu}^\rho 
\equiv \bar{\Gamma}_{\mu\nu}^\rho - \Gamma_{\mu\nu}^\rho 
= B \left[ A_{(\mu} F_{\nu)}\,^\rho
+ \frac{A^\rho}{1 - 2 B X} \left( \nabla_{(\mu} A_{\nu)} 
+ B A_\lambda A_{(\mu} \nabla^\lambda A_{\nu)}
+ B A_{(\mu} \nabla_{\nu)} X \right) \right] \; .
\label{delGamma-const}
\end{equation}
Here $V_{(\mu} W_{\nu)} \equiv ( V_\mu W_\nu + V_\nu W_\mu ) / 2$ is the symmetric part, and we define the antisymmetric part by square brackets $[ ...  ]$ in a similar manner.
Note that while $\Gamma_{\mu\nu}^\rho$ and $\bar{\Gamma}_{\mu\nu}^\rho$ are not individually tensorial, their difference $\delta\Gamma_{\mu\nu}^\rho$ is a tensor. The main difference in the transformation with respect to the scalar field case comes from the new term containing $F_{\mu\nu}$ which vanishes for $A_{\mu}=\nabla_{\mu}\phi$. We also use the well-known properties of the Levi-Civita tensor repeatedly, that is
\begin{equation}
\nabla_\lambda \epsilon^{\mu\nu\rho\sigma} = 0 \; , \quad
\epsilon^{\mu\nu\rho\sigma} \epsilon_{\alpha\beta\gamma\delta} = 
- 4! \, \delta^{[\mu}_\alpha \delta^\nu_\beta \delta^\rho_\gamma \delta^{\sigma]}_\delta \; ,
\label{epsilon-prop}
\end{equation}
and possible contractions of the latter. The minus sign in the second equation of \eqref{epsilon-prop} is due to our choice of Lorentzian signature of the metric.
In the upcoming subsections, we use the above transformation rules to compute each part of the action \eqref{eq:genProc} one by one.

\subsection{Transformation of $\bar S_2$ and $\bar S_3$}
\label{subsec:L2L3}

The transformation of $\bar S_2$ is trivial, i.e.
\begin{equation}
\bar S_2 = \int d^4x \sqrt{- \bar g} \, \bar G_2 \left( \bar X , \bar F , \bar U , \bar Y \right) = 
\int d^4x \sqrt{- g} \, G_2^* \left( X , F , U , Y \right) \; ,
\label{S2_transform}
\end{equation}
where
\begin{equation}
G_2^* \left( X, F, U, Y \right) \equiv \Omega^4 \sqrt{1 - 2 B X} \,
\bar{G}_2 \; ,
\label{def_G2}
\end{equation}
with the arguments of $\bar G_2 \left( \bar X , \bar F , \bar U , \bar Y \right)$ expressed by the ``unbarred'' quantities as in \eqref{XFUY}. The notation for $G_2^*$ will be clear later on.

The $\bar S_3$ term also transforms straightforwardly, that is
\begin{equation}
\bar S_3 = \int d^4x \sqrt{ - \bar g} \, \bar G_3 ( \bar X ) \bar\nabla_\mu \bar A^\mu = 
\int d^4x \sqrt{ - g} \, G_3 ( X ) \nabla_\mu A^\mu \; ,
\label{S3_transform}
\end{equation}
up to surface terms, where
\begin{equation}
G_3(X) \equiv \Omega^2 \int^{\frac{X}{\Omega^2 ( 1- 2 BX )}} d\bar{X}' \sqrt{ 1 + 2 \Omega^2 B \bar{X}'} \, \bar{G}_{3,\bar{X}} (\bar{X}')=
\int^X dX' \frac{\Omega^2}{\sqrt{1 - 2 B X'}} \, \partial_{X'} \bar{G}_3 \left( \frac{X'}{\Omega^2 (1 - 2 B X')} \right) \; .
\label{def_G3}
\end{equation}
From \eqref{S2_transform} and \eqref{S3_transform}, we observe that $S_2$ and $S_3$ are independently closed under the disformal transformation.

\subsection{Transformation of $\bar{S}_4$}
\label{subsec:L4}

So far $\bar S_2$ and $\bar S_3$ contained gravity only trivially, and the transformation was no different from the one on a flat geometry. However, non-minimal couplings between the vector field and the metric enter in the remaining $\bar S_{4,5,6}$, which makes the calculation of their transformation somewhat cumbersome. One simplification that may help us is to make full use of the antisymmetric properties of the Levi-Civita tensor. In this regard, we rewrite $\bar{\cal L}_4$ \eqref{ProcL4} as
\begin{equation}
\bar{\cal L}_4 = - \frac{1}{4} \, \bar\epsilon^{\alpha\beta\mu\nu} \bar\epsilon_{\alpha\beta\rho\sigma} \left[ \bar{G}_4 \big( \bar X \big) \bar{R}^{\rho\sigma}{}_{\mu\nu} 
+ 2 \bar{G}_{4,\bar{X}} \big( \bar X \big) 
\bar\nabla_\mu \bar{A}^\rho \bar\nabla_\nu \bar{A}^\sigma \right]
\; , 
\label{L4epsilon}
\end{equation}
where contraction of the Levi-Civita tensors is given by \eqref{epsilon-prop}. 
Using the transformations of $\bar\nabla_\mu \bar A_\nu$ and $\bar{R}^\mu{}_{\nu\rho\sigma}$ in \eqref{DA_Riemann} with $\delta\Gamma^\rho_{\mu\nu}$ in \eqref{delGamma-const}, we find, after some algebra, that the first term in \eqref{L4epsilon} transform as
\begin{equation}
\begin{aligned}
\int d^4x & \sqrt{-\bar{g}} \, \bar\epsilon^{\alpha\beta\mu\nu} \bar\epsilon_{\alpha\beta\rho\sigma} \left[ - \frac{1}{4} \, \bar{G}_4 \big( \bar X \big) \bar{R}^{\rho\sigma}{}_{\mu\nu} \right] \\
&
=  
\int d^4x \sqrt{-g} \, \Omega^2 \sqrt{1 - 2 B X} \, \bar{G}_4 \Bigg[ R
- \frac{B}{1 - 2 B X} \left( (\nabla_\mu A^\mu )^2 - \nabla_\mu A^\nu \nabla_\nu A^\mu \right) + \frac{B^2}{2 ( 1 - 2 B X)} \left( Y - 4 XF \right) \Bigg] \\
& \quad
- 2 \int d^4x \sqrt{1 - 2 B X} \, \frac{B \, \bar{G}_{4, \bar{X}}}{( 1 - 2 B X)^3} \nabla_\mu X 
\left( \nabla^\mu X + A^\mu \nabla_\nu A^\nu \right) \; ,
\end{aligned}
\label{G4calc}
\end{equation}
up to surface terms, and the second term in \eqref{L4epsilon} as
\begin{equation}
\begin{aligned} 
\int d^4x & \sqrt{- \bar{g}} \, \bar{G}_{4,\bar{X}} \, \bar\epsilon^{\alpha\beta\mu\nu} \bar\epsilon_{\alpha\beta\rho\sigma}
\bar\nabla_\mu \bar{A}^\rho \bar\nabla_\nu \bar{A}^\sigma \\
& =
 \int d^4x \sqrt{- g} \sqrt{1 - 2 B X} \, \bar{G}_{4,\bar{X}} \bigg[
\frac{1}{(1 - 2 B X)^2} \left( \left( \nabla_\mu A^\mu \right)^2 - \nabla_\mu A^\nu \nabla_\nu A^\mu \right)
- \frac{B (1 - B X)}{( 1 - 2 B X)^2} \left( Y - 4 XF \right) \\
& \qquad\qquad\qquad\qquad\qquad\qquad\qquad
+ \frac{2 B}{( 1 - 2 B X)^3} \nabla_\mu X \left( \nabla^\mu X + A^\mu \nabla_\nu A^\nu \right)
\bigg] \; .
\end{aligned}
\label{G4Xcalc}
\end{equation}
Notice that the last lines of \eqref{G4calc} and \eqref{G4Xcalc} conveniently cancel out each other. From \eqref{G4calc}, it is clear that we should define the new $G_4$ as
\begin{equation}
G_4(X) \equiv \Omega^2 \sqrt{1 - 2 B X} \, \bar G_4 \; ,
\end{equation}
where the argument of $\bar G_4 (\bar X)$ is expressed by the ``unbarred'' $X$ as in \eqref{XFUY}. Adding \eqref{G4calc} and \eqref{G4Xcalc}, we obtain the transformed $\bar S_4$ as
\begin{eqnarray}
 \bar{S}_4 & = & \int d^4x \sqrt{ - \bar{g}} \, \bar{\cal L}_4 \nonumber\\
 & = & 
  \int d^4x \sqrt{- g} \Bigg[
  G_{4} R + G_{4,X} 
  \left( ( \nabla_\mu A^\mu )^2 - \nabla_\mu A^\nu \nabla_\nu A^\mu \right) 
  - 
  \left( \frac{1}{2} \, G_4 B^2 + G_{4,X} B \left( 1  - B X \right) \right) \left( Y - 4 XF \right) 
  \Bigg]\,,
\label{S4trans}
\end{eqnarray}
up to surface terms. We see that, unlike $\bar S_2$ and $\bar S_3$, the part $\bar{S}_4$ transforms into $S_4$ plus additional terms that contribute to $S_2$ and thus is not closed under the disformal transformation by itself. This feature is absent in the Horndeski theory, where $G_4$ is closed under a constant disformal transformation (see for example Refs.~\cite{Bettoni,BenAchour:2016fzp,Crisostomi:2016tcp}). The extra terms in \eqref{S4trans} indeed vanish by substituting $A_\mu = \nabla_\mu \phi$, which leads to $F_{\mu\nu} =0$ and thus to $F = Y = 0$.

An interesting case to consider is the transformation of the Einstein-Hilbert (EH) action, which corresponds to fixing $\bar{G}_4 = M_{\rm Pl}^2 / 2$, where $M_{\rm Pl}$ is the reduced Planck mass. We obtain the following Lagrangian in the disformal frame (i.e.~the ``unbarred'' coordinates),
\begin{equation}
\bar S_{\rm EH} 
= \frac{M_{\rm Pl}^2}{2} \int d^4x \sqrt{-g} \,
\left[ \Omega^2 \sqrt{1 - 2 B X} \, R - \frac{\Omega^2 B}{\sqrt{1 - 2 B X}} \left( (\nabla_\mu A^\mu)^2 - \nabla_\rho A^\sigma \nabla_\sigma A^\rho \right) 
+ \frac{\Omega^2 B^2}{2 \sqrt{ 1 - 2 B X}} \left( Y - 4 X F \right) \right] \; .
\label{eq:hilbertransf}
\end{equation}
This is a special case of Jordan-frame action of a non-canonical vector field. 
At a first glance, the action \eqref{eq:hilbertransf} appears to contain a vector field non-minimally coupled to gravity and thus to retain total five physical degrees of freedom. However, since it is related to the EH action with an invertible transformation, the former must be equivalent to the latter. Similarly to the case of scalar-dependent transformation \cite{Domenech:2015tca}, the action \eqref{eq:hilbertransf} is in fact a highly constrained system and propagates only two degrees of freedom as in the EH theory.

\subsection{Transformation of $\bar S_5$}
\label{subsec:L5}

Similarly to the computation in the previous subsection, it is useful to use the properties of the Levi-Civita tensor in finding the transformation rules of $\bar S_5$. We can express $\bar{\cal L}_5$ of the form \eqref{ProcL5} as
\begin{eqnarray}
\bar{\cal L}_5 & = &
\bar{\cal L}_{G_5} + \bar{\cal L}_{g_5} + \bar{\cal L}_{{\cal G}_5} \; , 
\label{L5epsilon} \\
& \bar{\cal L}_{G_5} & \equiv 
\frac{1}{4} \, \bar\epsilon^{\lambda\alpha\mu\nu} \bar\epsilon_{\lambda\beta\rho\sigma} \bar{\nabla}_\alpha \bar{A}^\beta 
\left[ \bar{G}_5 ( \bar X ) \,  \bar{R}^{\rho\sigma}{}_{\mu\nu}
+ \frac{2}{3} \bar{G}_{5, \bar{X}} ( \bar X )
\bar\nabla_\mu \bar{A}^\rho \bar\nabla_\nu \bar{A}^\sigma
\right] \; , \\
& \bar{\cal L}_{g_5} & \equiv
{-} \bar{g}_5(\bar X) \, \bar\epsilon^{\lambda\alpha\mu\nu} \bar\epsilon_{\lambda\beta\rho\sigma} \bar{\nabla}_\alpha \bar{A}^\beta \bar\nabla_\mu \bar{A}_\nu \bar\nabla^\rho \bar{A}^\sigma \; ,\\
& \bar{\cal L}_{{\cal G}_5} & \equiv
{-} \bar{\cal G}_5 ( \bar{X} ) \, \bar\epsilon^{\mu\alpha\rho\sigma} \bar\epsilon_{\nu\beta\lambda\tau} 
\bar{A}_\mu \bar{A}^\nu 
\bar\nabla_{\rho} \bar{A}_\sigma \bar\nabla^\lambda \bar{A}^\tau
\bar{\nabla}_\alpha \bar{A}^\beta \label{eq:newG5}\; .
\end{eqnarray}
Using the transformation laws of various quantities summarized in \eqref{ginv_det}--\eqref{delGamma-const}, we can compute how these terms transform. Keeping the Levi-Civita tensors in \eqref{L5epsilon} until the very last step of the computation is somewhat helpful to keep track of a large number of terms.
Note that $\bar{\cal L}_{g_5}$ and $\bar{\cal L}_{{\cal G}_5}$ are identically null if we inject $\bar A_\mu = \bar\nabla_\mu\phi$ and thus they are terms that are novel in comparison to the Horndeski theory. Since $\bar G_5$, $\bar g_5$ and $\bar{\cal G}_5$ are independent functions, we will treat the transformation of each of $\bar{\cal L}_{G_5}$, $\bar{\cal L}_{g_5}$ and $\bar{\cal L}_{{\cal G}_5}$ separately.

Let us first consider the transformation of $\bar{\cal L}_{G_5}$. This invokes the most lengthy calculation among the terms in $\bar{\cal L}_5$ due to the presence of the Riemann/Einstein tensor. As in the transformation of the $\bar G_4$ terms in the previous section, there are non-trivial (and more involved) cancellations between the terms from the transformation of the $\bar G_5$ term and those from $\bar G_{5,\bar X}$.
After a straightforward computation, one obtains

\begin{equation}
\begin{aligned}
\int d^4x \sqrt{- \bar g} \, \bar{\cal L}_{G_5} = 
\int d^4x \sqrt{-g}\bigg[ &
G_5(X) \, G_{\mu\nu} \nabla^\mu A^\nu 
- \frac{1}{6} \, G_{5, X}(X)
\left(
\left( \nabla_\mu A^\mu \right)^3 - 3 \nabla_\mu A^\nu \nabla_\nu A^\mu \nabla_\rho A^\rho + 2 \nabla_\mu A^\nu \nabla_\rho A^\mu \nabla_\nu A^\rho
\right) \\
&
+ \frac{1}{2} \, B  \left(1 - B X \right) G_{5,X}(X) \, \tilde F^{\mu\rho} \tilde F_{\nu\rho} A_\mu A^\nu \nabla_\rho A^\sigma  \bigg] \; ,
\end{aligned}
\label{LG5_transform}
\end{equation}
up to total derivatives, where
\begin{equation}
G_5(X) \equiv \int^{\frac{X}{\Omega^2 ( 1 - 2 BX )}} d\bar{X}' \sqrt{1 + 2 \Omega^2 B \bar{X}'} \, \bar{G}_{5,\bar{X}} (\bar{X}') = \int^X \frac{dX'}{\sqrt{1 - 2 B X'}} \, \partial_{X'} \bar{G}_{5} \left( \frac{X'}{\Omega^2 ( 1 - 2 B X')} \right) \; .
\end{equation}
The first line in \eqref{LG5_transform} is indeed the corresponding ${\cal L}_{G_5}$ term in the transformed action. Besides, the second line in \eqref{LG5_transform} contributes to ${\cal L}_{{\cal G}_5}$ in the new action. This is the very reason why ${\cal L}_{{\cal G}_5}$ is necessary for the ${\cal L}_5$ part of the action to be closed under the considered disformal transformation.

The transformation of remaining terms, $\bar{\cal L}_{g_5}$ and $\bar{\cal L}_{{\cal G}_5}$, is much simpler. One can straightforwardly find, for $\bar{\cal L}_{g_5}$, 
\begin{equation}
\sqrt{- \bar g} \, \bar{\cal L}_{g_5}
= {-} \sqrt{-g} \left[ g_5(X) \tilde F^{\mu\rho} \tilde F_{\nu\rho} \nabla_\mu A^\nu 
+ B g_5(X) \tilde F^{\mu\rho} \tilde F_{\nu\sigma} A_\mu A^\nu \nabla_\rho A^\sigma \right]\,,
\label{Lg5_transform}
\end{equation}
where
\begin{equation}
g_5(X) \equiv \frac{\bar{g}_5}{\Omega^2 ( 1 - 2 B X)^{3/2}} \; ,
\end{equation}
with the argument of $\bar{g}_5(\bar X)$ expressed by the ``unbarred'' $X$ as in \eqref{XFUY}.
Eq.~\eqref{Lg5_transform} tells us that the transformation of the $\bar{\cal L}_{g_5}$ term indeed produces the corresponding ${\cal L}_{g_5}$ in the new action, with an additional contribution to ${\cal L}_{{\cal G}_5}$, which is the last term in \eqref{Lg5_transform}. Thus the presence of ${\cal L}_{{\cal G}_5}$ is again necessary for the closure under the transformation.
On the other hand, the $\bar{\cal L}_{{\cal G}_5}$ term is by itself closed under the considered transformation, namely,
\begin{equation}
\sqrt{- \bar{g}} \, \bar{\cal L}_{{\cal G}_5}
= {-} \sqrt{-g} \, {\cal G}^*_5(X) \tilde F^{\mu\rho} \tilde F_{\nu\sigma} A_\mu A^\nu \nabla_\rho A^\sigma \; ,
\label{LcalG_transform}
\end{equation}
where
\begin{equation}
{\cal G}^*_5 \equiv \frac{\bar{\cal G}_5}{\Omega^4 ( 1 - 2 B X)^{3/2}} \; ,
\end{equation}
with, again, the argument of $\bar{\cal G}_5(\bar X)$ expressed in terms of the ``unbarred'' $X$. The notation ${\cal G}^*_5 $ will be clear later.  

Combining \eqref{LG5_transform}, \eqref{Lg5_transform} and \eqref{LcalG_transform}, we observe that, despite the fact that each term in $\bar S_5$ is not on its own closed under the transformation (except for ${\cal L}_{{\cal G}_5}$), $\bar S_5$ as a whole is indeed closed. Overall, the $\bar{S}_5$ transforms as
\begin{equation}
\bar{S}_5 = S_5 \; ,
\end{equation}
up to total derivatives, with the redefinition of the functions
\begin{equation}
\begin{aligned} 
G_5(X) & \equiv \int^X dX' \frac{\partial_{X'} \bar{G}_5}{\sqrt{1 - 2 B X'}} \; , \quad
g_5(X) \equiv \frac{\bar g_5}{\Omega^2 ( 1 - 2 B X)^{3/2}} \; , \quad
{\cal G}_5(X) \equiv {\cal G}_5^* 
{-} \frac{B \left(1 - B X \right)}{2} \, G_{5,X} + B g_5 \; .
\end{aligned}
\end{equation}
We emphasize again that the presence of the $\bar{\cal G}_5$ term is mandatory so that $\bar S_5$ is closed under the transformation.

\subsection{Transformation of $\bar S_6$}
\label{subsec:L6}

Let us now move on to the last piece of our entire action. As in the previous two subsections for $\bar S_4$ and $\bar S_5$, we rewrite $\bar{\cal L}_6$ using the Levi-Civita tensors as
\begin{equation}
\bar{\cal L}_6 = \frac{1}{4} \, \bar{\epsilon}^{\mu\nu\alpha\beta} \bar{\epsilon}_{\rho\sigma\gamma\delta} \left[ \bar{G}_6 \big( \bar X \big) \bar{R}_{\alpha\beta}{}^{\gamma\delta} \bar{\nabla}_\mu \bar{A}_\nu  \bar{\nabla}^\rho \bar{A}^\sigma
+ 2 \bar{G}_{6,\bar{X}} \big( \bar X \big)
\bar\nabla_\alpha \bar{A}_\beta \bar\nabla^\gamma \bar{A}^\delta
\bar{\nabla}_\mu \bar{A}^\rho \bar{\nabla}_\nu \bar{A}^\sigma 
\right] \; .
\label{L6epsilon}
\end{equation}
In order to remove unwanted second-order derivatives from the action after plugging in \eqref{disformal}, it is advisable to use $\bar\nabla_\mu$ rather than $\nabla_\mu$ to take away total derivatives. Then, with the use of the Bianchi identity $\nabla_{[\mu} F_{\nu\rho]} = 0$, one can entirely remove such unwanted terms and easily infer the form of the new function $G_6$. Heavily using the properties of the Levi-Civita tensor, and making use of the identity relations \cite{Fleury:2014qfa}
\begin{equation}
F_{\mu\rho} F^{\nu\rho} - \tilde F_{\mu\rho} \tilde F^{\nu\rho} = \frac{1}{2} \, F_{\rho\sigma} F^{\rho\sigma} \, \delta^\nu_\mu
= - 2 F \, \delta^\nu_\mu \; , \quad
F_{\mu\rho} \tilde F^{\nu \rho} = \frac{1}{4} \, F_{\rho\sigma} \tilde F^{\rho\sigma} \, \delta^\nu_\mu 
= - U \, \delta^\nu_\mu \; ,
\end{equation}
we obtain in the end
\begin{equation}
\begin{aligned}
\bar S_6 = \int d^4x \sqrt{- \bar g} \, \bar{\cal L}_6 = 
\int d^4x \sqrt{-g} \bigg[ &
G_6(X) L^{\mu\nu\rho\sigma} \nabla_{\mu}A_\nu \nabla_{\rho}A_\sigma 
+ \frac{1}{2} \, G_{6,X}(X) \tilde{F}^{\mu\rho} \tilde{F}_{\nu\sigma} \nabla_\mu A^\nu \nabla_\rho A^\sigma \\
& - \left(
\frac{1}{2} \, G_6 \left( 2 - B X \right) 
+ G_{6,X} X \left( 1 - B X \right)
\right) B U^2 \bigg] \; ,
\end{aligned}
\label{L6_transform}
\end{equation}
up to surface terms, where
\begin{equation}
G_6 \equiv \frac{\bar{G}_6}{\Omega^2 \sqrt{ 1 - 2 B X}} \; ,
\end{equation}
with the argument of $\bar G_6 (\bar X)$ expressed in terms of the ``unbarred'' $X$ using \eqref{XFUY}. During the computation, there are a large number of cancellations between the terms coming from the transformation of the $\bar G_6$ term and those from $\bar G_{6,\bar X}$, similarly to the cases of $\bar S_4$ and $\bar S_5$ in the previous subsections.
As is seen in \eqref{L6_transform}, $\bar S_6$ is not closed under the considered disformal transformation and it gives rise to contribution to the $S_2$ term in the new action, which is depicted in the second line of \eqref{L6_transform}.

\subsection{Summary of transformation}
\label{subsec:trans_summary}

To conclude this section, we would like to summarize the results obtained in the previous subsections for the disformal transformation \eqref{disformal} on the generalized Proca action \eqref{eq:genProc}. Each term $\bar S_{2,3,4,5,6}$ transforms as
\begin{subequations}
\label{S2toS6}
\begin{align}
\bar S_2 & = \int d^4x \sqrt{- g} \, \Omega^4 \sqrt{1 - 2 B X} \, \bar{G}_2 \; , \\
\bar S_3 & = S_3 \; , \\
\bar S_4 & = S_4 
- \int d^4x \sqrt{-g} 
\left[ \frac{1}{2} \, G_4 B^2 + G_{4,X} B \left( 1  - B X \right) \right] \left( Y - 4 XF \right) \; , \\
\bar S_5 & = S_5 \; , \\
\bar S_6 & = S_6 
- \int d^4x \sqrt{-g}
\left[
\frac{1}{2} \, G_6 \left( 2 - B X \right) 
+ G_{6,X} X \left( 1 - B X \right)
\right] B U^2 \; ,
\end{align}
\end{subequations}
up to surface terms, where the new, transformed $S_{3,4,5,6}$ take the same form as in \eqref{ProcL3}--\eqref{ProcL6} with redefined functions $G_{3,4,5,6}$, $g_5$ and ${\cal G}_5$ that are related to the original ones $\bar G_{3,4,5,6}$, $\bar g_5$ and $\bar{\cal G}_5$ as
\begin{subequations}
\label{G3toG6}
\begin{align}
G_3(X) & \equiv
\int^X dX' \frac{\Omega^2}{\sqrt{1 - 2 B X'}} \, \partial_{X'} \bar{G}_3 \left( \frac{X'}{\Omega^2 (1 - 2 B X')} \right) \; , \\
G_4(X) & \equiv \Omega^2 \sqrt{1 - 2 B X} \, \bar G_4 \; , \\
G_5(X) & \equiv \int^X \frac{dX'}{\sqrt{1 - 2 B X'}} \, \partial_{X'} \bar{G}_{5} \left( \frac{X'}{\Omega^2 ( 1 - 2 B X')} \right) \; , \\
g_5(X) & \equiv \frac{\bar{g}_5}{\Omega^2 ( 1 - 2 B X)^{3/2}} \; , \\
{\cal G}_5(X) & \equiv \frac{\bar{\cal G}_5}{\Omega^4 ( 1 - 2 B X)^{3/2}} 
{-} \frac{B \left(1 - B X \right)}{2} \, G_{5,X} + B g_5 \; , \\
G_6(X) & \equiv \frac{\bar{G}_6}{\Omega^2 \sqrt{ 1 - 2 B X}} \; .
\end{align}
\end{subequations}
Note that the arguments of $\bar G_{2,3,4,5,6}$, $\bar g_5$ and $\bar{\cal G}_5$ appearing on the right-hand sides of \eqref{S2toS6} and \eqref{G3toG6} are now all written in terms of ``unbarred'' quantities as in (\ref{XFUY}).
Compiling the contributions from $\bar S_4$ and $\bar S_6$ to the new $S_2$, we now define the new function $G_2$ as
\begin{equation}
\begin{aligned}
G_2(X,F,U,Y) \equiv \, & \Omega^4 \sqrt{1 - 2 B X} \, \bar{G}_2
- \left[ \frac{1}{2} \, G_4 B^2 + G_{4,X} B \left( 1  - B X \right) \right] \left( Y - 4 XF \right) \\ & 
- \left[
\frac{1}{2} \, G_6 \left( 2 - B X \right) 
+ G_{6,X} X \left( 1 - B X \right)
\right] B U^2 \; .
\end{aligned}
\label{G2_transform}
\end{equation}
Therefore the entire generalized Proca action transforms as
\begin{equation}
\bar S_{\rm GP} = \sum_{i=2}^6 \bar S_i = \sum_{i=2}^6 S_i = S_{\rm GP} \; ,
\end{equation}
\begin{figure}
 \includegraphics[width=0.5\columnwidth]{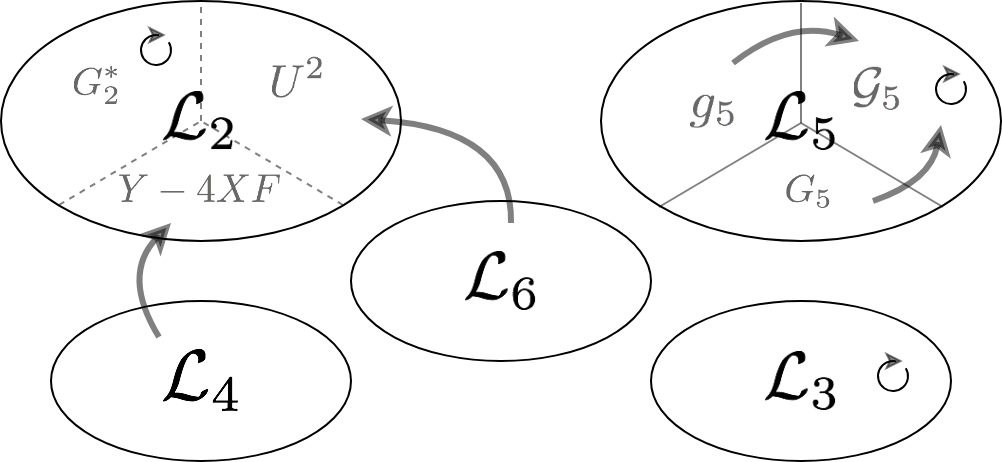}
\caption{ \label{fig:diagram}
Schematic diagram of the transformations of the generalized Proca Lagrangian. An arrow indicates that after a disformal transformation extra terms will contribute to the Lagrangian to which the arrow is directed. A loop means that the Lagrangian is closed by itself under a disformal transformation. We have to include ${\cal G}_5$ and $U^2$ terms so that the generalized Proca Lagrangian is closed. The division of ${\cal L}_2$ is only for illustrative purposes and to visualize the contributions from ${\cal L}_4$ and ${\cal L}_6$ to parts of ${\cal L}_2$ through the transformation. See the main text for details.}
\end{figure}
up to surface terms, with replacement of the arbitrary functions defined in \eqref{G3toG6} and \eqref{G2_transform}. The generalized Proca theory \eqref{eq:genProc} is thus closed under the disformal transformation \eqref{disformal}. This closure is, however, valid only when we impose the constancy of the conformal and disformal factors $\Omega$ and $B$, and requires the presence of all $X,F,U,Y$ dependences of $\bar G_2$ and the presence of the $\bar{\cal G}_5$ term in $\bar S_5$. See FIG.~\ref{fig:diagram} for a schematic picture of the closure structure. In the following section, based on the results so far, we provide an extended analysis of the transformation of sound speeds of perturbations around cosmological backgrounds.

\section{Sound speeds and metric transformation}
\label{sec:soundspeeds}

Disformal transformations with scalar fields become simple enough in the uniform scalar slicing. The generalization to the case of vector fields is similar but not straightforward; there is no equivalent to the uniform scalar slicing. Nevertheless, it has been shown that for a timelike vector, gauge invariant variables are invariant under a disformal transformation and, hence, a disformal transformation does not change the definition of observables \cite{Papadopoulos:2017xxx}. 

On the other hand, it is well known that disformal couplings to matter change the relative propagation speed of the fields, e.g.~between photons and GWs, on cosmological backgrounds. In the scalar field case, a disformal transformation only affects the 00 component of the Friedmann-Lema\^{i}tre-Robertson-Walker (FLRW) metric and so we expect that the sound speed gets rescaled simply due to the rescaling of time \cite{Domenech:2015hka}. Whether the same logic applies to a transformation with vector fields requires an explicit check, since there is a non-trivial mixing of the perturbation variables. In the following subsections we will review the cosmological perturbations in a FLRW background and we will explicitly show that sound speeds transform as expected, that is, simply interpreted as changing the background lightcone structure.

\subsection{FLRW background and sound speeds}
Let us briefly review the results derived in Ref.~\cite{DeFelice:2016uil,DeFelice:2016yws}. 
Considering the flat FLRW background metric
\begin{equation}
ds^2 = - dt^2 + a^2(t) \, \delta_{ij} \, dx^i dx^j \; ,
\label{FLRWmetric}
\end{equation}
and taking the background vacuum expectation value of the vector field $A_\mu$ compatible with \eqref{FLRWmetric} as~%
\footnote{Without additional matter, background solutions fix $\phi = {\rm const.}$ and $H = {\rm const.}$, where $H \equiv \dot{a} / a$ is the Hubble parameter.}
\begin{equation}
\left\langle A_\mu \right\rangle = - \phi(t) \, \delta_\mu^0 \; ,
\label{vectorvev}
\end{equation}
we can expand the perturbations of the metric and the vector field around the homogeneous and isotropic background given by \eqref{FLRWmetric} and \eqref{vectorvev}. These perturbation degrees of freedom can be decomposed into scalar, vector and tensor sectors that are mutually decoupled from each other at the first order in perturbative expansion.
The system of perturbations contains three degrees of freedom (dof) in addition to the standard two in gravity, in total five dof. While the scalar sector consists of only one dof, each of the vector and tensor sectors has two dof. Both tensor modes follow exactly the same equations of motion. Both vector modes have the same kinetic coefficient and propagation speed in the large momentum limit only. Thus, for our purpose, it is sufficient to consider only one formula of the sound speed for each sector.

After the reduction of the system, the kinetic coefficient ($q_T$) and the squared sound speed ($c_T^2$) of tensor modes are respectively given by
\begin{align}\label{eq:tensors}
q_T=2G_4-2\phi^2G_{4,X}+H\phi^3 G_{5,X}
\quad{\rm and}\quad
c^2_T=\frac{2G_4+\phi^2\dot\phi G_{5,X}}{q_T}\,.
\end{align}
For the vector modes we have
\begin{align}\label{eq:vectors}
q_V=G_{2,F}+2G_{2,Y}\phi^2-4g_5H\phi+2G_6H^2+2G_{6,X}H^2\phi^2
\quad{\rm and}\quad
c_V^2=\frac{\mu_V}{q_V}\,,
\end{align}
where
\begin{align}
\mu_V=G_{2,F}+2G_6\left(\dot H+H^2\right)+2\dot\phi\phi H G_{6,X}-2g_5\left(H\phi+\dot\phi\right)+2{\cal G}_5\phi^3H+\frac{\phi^2\left(2G_{4,X}-G_{5,X}\phi H\right)^2}{2q_T}\,.
\label{muV}
\end{align}
Note that we have included the effects of the new term ${\cal G}_5$, as compared to the expressions in \cite{DeFelice:2016yws}.
The scalar sector is more involved. Some useful quantities to define for shorthand notation are given by
\begin{align}
w_1&=H^2\phi^3\left(G_{5,X}+\phi^2G_{5,XX}\right)-4H\left(G_4+\phi^4G_{4,XX}\right)-\phi^3 G_{3,X}\\
w_2&=w_1+2Hq_T\\
w_3&=-2\phi^2 q_V\\
w_4&=\frac{1}{2}H^3\phi^3\left(9G_{5,X}-\phi^4G_{5,XXX}\right)-3H^2  \left(2G_4+2\phi^2G_{4,X}+\phi^4G_{4,XX}-\phi^6G_{4,XXX} \right) \nonumber \\&\quad
-\frac{3}{2}H\phi^3\left(G_{3,X}-\phi^2G_{3,XX}\right)+\frac{1}{2}\phi^4G_{2,XX}\\
w_5&=w_4-\frac{3}{2}H\left(w_1+w_2\right)\\
w_6&=-\phi\left[H^2\phi\left(G_{5,X}-\phi^2G_{5,XX}\right)-4H\left(G_{4,X}-\phi^2G_{4,XX}\right)+\phi G_{3,X}\right]\\
w_7&=2\left(H\phi G_{5,X}-2 G_{4,X}\right)\dot H+\left[H^2\left(G_{5,X}+\phi^2 G_{5,XX}\right)-4H\phi G_{4,XX}-G_{3,X}\right]\dot\phi\,.
\end{align}
Then the kinetic coefficient ($Q_S$) and the squared sound speed ($c_S^2$) of scalar perturbations are given by
\begin{align}
Q_S=\frac{H^2 q_Tq_S}{\phi^2\left(w_1-2w_2\right)^2} \quad{\rm and}\quad
 c_S^2 = \frac{\mu_S}{8H^2\phi^2q_Tq_Vq_S}\,,
\end{align}
where
\begin{align}
	q_S & \equiv 3w_1^2+4q_Tw_4 \; , \\
	\mu_S &\equiv\left[w_6\phi\left(w_1-2w_2\right)+w_1w_2\right]^2-w_3\left(2w_2^2\dot w_1-w_1^2\dot w_2\right)+\phi\left(w_1-2w_2\right)^2w_3\dot w_6 \nonumber \\&\quad
+w_3\left(w_1-2w_2\right)\left[\left(H-2\dot\phi/\phi\right)w_1w_2+\left(w_1-2w_2\right)\left\{w_6\left(H\phi-\dot\phi\right)+2w_7\phi^2\right\}\right]\,.
\end{align}
The stability of the system in the subhorizon limit is certified by the conditions $q_T>0$, $q_V>0$, $q_S>0$, $2 G_4 + \phi^2 \dot\phi G_{5,X} \ge 0$, $\mu_V \ge 0$ and $\mu_S \ge 0$.
The above formulas are valid in any frame, since they are calculated for a general functional form of the Generalized Proca action. In the original frame they will be given by the ``barred'' functions and in the new frame with the ``unbarred'' ones. In what follows we will start from the ``barred'' frame variables and rewrite them in terms of ``unbarred'' ones according to the transformation rules derived in Sec.~\ref{sec:Proca}. We neglect matter field for the moment and focus only in the change of the gravitational sector. We will address the coupling to matter fields at the end of the section. 

For later use we present here the transformation rules for background quantities. They are given by
\begin{align}
\bar a = \Omega a\quad ,\quad d\bar t=\Omega\sqrt{1-2 B X}\,dt\quad,\quad\bar H =\frac{H}{\Omega\sqrt{1-2 B X}}\quad ,\quad \bar\phi=\frac{\phi}{\Omega\sqrt{1-2 B X}} \; ,
\end{align}
and therefore we also have
\begin{align}
\frac{d\bar H}{d\bar t} =\frac{H}{\Omega^2\left(1-2 B X\right)}\left(\frac{\dot H}{H}+\frac{\dot\phi}{\phi}\frac{2B X}{1-2 B X}\right)\quad {\rm and}\quad 
\frac{d\bar\phi}{d\bar t} = \frac{\dot\phi}{\Omega^2 \left( 1 - 2 B X \right)^2}\,,
\end{align}
where $X = \phi^2 /2$, and $d\bar{t}$ and $\bar\phi$ shift due to the change of the lapse. (Recall that we only consider transformations with constant $\Omega$ and $B$.)
Also, when we need to find how derivatives of the functions, especially of $G_2$, transform, it is useful to know the following partial derivatives
\begin{equation}
\left(
\begin{array}{cccc}
\partial X / \partial \bar X & \partial X / \partial \bar F & \partial X / \partial \bar Y & \partial X / \partial \bar U \\
\partial F / \partial \bar X & \partial F / \partial \bar F & \partial F / \partial \bar Y & \partial F / \partial \bar U \\
\partial Y / \partial \bar X & \partial Y / \partial \bar F & \partial Y / \partial \bar Y & \partial Y / \partial \bar U \\
\partial U / \partial \bar X & \partial U / \partial \bar F & \partial U / \partial \bar Y & \partial U / \partial \bar U \\
\end{array}
\right) = \left(
\begin{array}{cccc}
\Omega^2 \left( 1 - 2 B X \right)^2 & 0 & 0 & 0 \\
\Omega^2 B^2 Y & \Omega^4 & - \frac{\Omega^6}{2} B \left( 1 - 2 B X \right) & 0 \\
- 4\Omega^2 B \left( 1 - 2 B X \right) Y & 0 & \Omega^6 \left( 1 - 2 B X \right)^2 & 0 \\
-\Omega^2 B \left( 1 - 2 B X \right) U & 0 & 0 & \Omega^4 \sqrt{1 - 2 B X} \\
\end{array}
\right) \; ,
\label{partialders}
\end{equation}
while $Y = U = 0$ on the background (\ref{FLRWmetric}, \ref{vectorvev}).
With these formulas we are ready to explicitly check the transformation rule of the sound speed for each sector.

\subsection{Transformation of $c_{\rm T}^2$}

The transformation rule for tensor modes is the simplest one. A short algebra with Eqs.~\eqref{eq:tensors} and \eqref{G3toG6} tells us that 
\begin{align}\label{eq:qtct}
\bar q_T=\Omega^{-2}\sqrt{1-2 B X}\, q_T
\quad
{\rm and}
\quad
\bar c^2_T=\frac{c_T^2}{1-2B X}\,.
\end{align}
\subsection{Transformation of $c_{\rm V}^2$}

The transformation of vector modes is not as straightforward as in the tensor case but we can easily see how terms compensate each other or how they do not contribute after the transformation. In this direction, it is important to note that (see \eqref{partialders})
\begin{align}
\bar G_{2,\bar F}=\Omega^4\bar G_{2, F}\quad{\rm and}\quad
\bar G_{2,\bar Y}=\Omega^6\left(1-2B X\right)^2\left[\bar G_{2,Y}-\frac{B/2}{1-2B X}\bar G_{2,F}\right]\,.
\end{align}
From this relation and \eqref{G2_transform} we immediately see that the combination that appears in $q_V$ just rescales as
\begin{align}
\bar G_{2,\bar F}+2\bar G_{2,\bar Y}\bar \phi^2= 
\sqrt{1-2B X} \left(G_{2,F}+2G_{2,Y}\phi^2\right)\,.
\end{align}
This also implies that the combination $\hat G_2\propto Y-4XF$ that, for example, appears in $G_2$ \eqref{G2_transform} from the transformation of $\bar G_4$ does not contribute to $q_V$ as $\hat G_{2,F}+2\hat G_{2,Y}\phi^2=0$. A similar cancellation takes place for the terms $\propto U^2$ coming from the transformation of $\bar G_6$. It is easy to check that the terms containing $\bar g_5$ and $\bar G_6$ scale in the same way and
\begin{align}
\bar q_V =\sqrt{1-2B X}\,q_V\,.
\end{align}
The calculation concerning $\mu_V$ is slightly more involved. First note that the terms containing $\bar G_6$ compensate by themselves. The terms containing $G_4$ and $G_5$ transform as
\begin{align}\label{eq:formuv}
2\bar G_{4,\bar X}-\bar G_{5,\bar X}\bar\phi \bar H=\sqrt{1-2B X}\left(2\ G_{4,X}- G_{5,X}\phi H+ B q_T\right)\,.
\end{align}
A short algebra shows that the extra terms coming from squaring Eq.~\eqref{eq:formuv} cancel with those coming from $\bar {\cal G}_5$ and the $G_4$ terms from $\bar G_2$ in $\mu_V$, 
see Eq.~\eqref{muV}. In the end, we find that
\begin{align}
\bar \mu_V = \frac{\mu _V}{\sqrt{1-2B X}}\quad {\rm and \quad thus}\quad
\bar c^2_V=\frac{c_V^2}{1-2B X}\,.
\end{align}

\subsection{Transformation of $c_{\rm S}^2$}
To understand the transformation rule for scalar perturbations it is sufficient to show that
\begin{align}
	\bar w_1&=\Omega^{-3}\left[w_1+2B X\left(w_1-2w_2\right)\right] \; , \\
	\bar w_2&=\Omega^{-3}\left[w_2+2B X\left(w_1-2w_2\right)\right] \; , \\
	\bar w_3&= \frac{w_3}{\Omega^2 \sqrt{ 1 - 2 B X}} \; ,  \\
	\bar w_4&=\Omega^{-4}\left[\left(1-2B X\right)^{3/2}w_4 +12 B X Hw_1\sqrt{1-2B X}-\frac{48H^2B^2 X^2 q_T}{\sqrt{1-2B X}}\right] \; , \\
	\bar w_6&=\frac{1}{\Omega^2\sqrt{1-2B X}}\left[w_6-\phi B\left(w_1+w_2 + 2 B X \left(w_1-2w_2\right)\right)\right] \; , \\
	\bar w_7&=\frac{1}{\Omega^2\sqrt{1-2B X}}\left(w_7-2B q_T\dot H-\frac{\dot\phi}{\phi} B w_2\right) \; , 
\end{align}
where we did not include $w_5$ as its transformation rule follows directly from those for $w_1$, $w_2$ and $w_4$. We have also used the background equations of motion to obtain the form of $\bar w_4$. It is important to note that $\rho_m+P_m=-2q_T\dot H- w_2 \dot\phi/\phi$, and therefore in absence of matter $\bar w_7$ just gets rescaled. With these transformation rules we find that
\begin{align}
 \bar q_S =\Omega^{-6}\left(1-2B X\right)^2 q_S\quad,\quad
 \bar Q_S =\Omega^{-2}\sqrt{1-2B X} \, Q_S\quad,\quad
 \bar \mu_S =\Omega^{-12}\mu_S\quad{\rm and}\quad
 \bar c_S^2 =\frac{c_S^2}{1-2B X}\,.
\end{align}

\subsection{Confirmation of expectation}

The previous results on the transformation rules of sound speeds were to be expected if we look at the equations of motion of the perturbation variables. The tensor modes follow a wave-like equation in the large momentum $k$ limit, which transform under the present disformal transformation as
\begin{align}
	\frac{1}{\bar a^3 \bar q_T}\frac{d}{d\bar t}\left(\bar a^3 \,\bar q_T\frac{d h_\lambda}{d\bar t}\right)+\bar c_T^2\frac{k^2}{\bar a^2} h_\lambda =\frac{1}{\Omega^{2}\left(1-2B X\right)}\left[\frac{1}{a^3 q_T}\frac{d}{dt}\left( a^3 \, q_T\frac{d h_\lambda}{d t}\right)+\left(1-2B X\right)\bar c_T^2\frac{k^2}{a^2} h_\lambda \right]=0\, .
\end{align}	
Completely analogous transformations take place for the scalar and vector modes.
From this wave equation we see that first the conformal factor does not affect the sound speed (as it does not change the light cone) but a disformal transformation does, with the factor given by Eq.~\eqref{eq:qtct}. Note that once the action of the whole system including the matter sector and its coupling to gravity is specified, this transformation does not change physical observables; the ``change'' in the sound speed is rather an effect of working with different variables \cite{Domenech:2015hka}. What is observable is e.g.~the ratio $c_{S,V,T}^2/c_{\gamma}^2$, which is invariant under the disformal transformation. On the other hand, if the matter sector is minimally coupled to the metric after the transformation then the ratio $c_{S,V,T}^2/c_{\gamma}^2$ is different from that for the case where the matter is coupled to the metric before the transformation. Alternatively, if two different matter components gravitate through the two different and disformally related metrics, $g_{\mu\nu}$ and $\bar g_{\mu\nu}$, then the sound speed is effectively measured in two different frames, and the difference in $c_T^2$ becomes a physically meaningful quantity.

\section{Stronger constraint on fine-tuned theory due to inhomogeneities}
\label{sec:constraint}

After the events GW170817~\cite{TheLIGOScientific:2017qsa} and GRB170817A~\cite{Monitor:2017mdv}, the theoretical space for the Horndeski and generalized Proca theories has been substantially narrowed down. So far, only models without $X$ dependence on the functions $G_4$ and $G_5$ or those with very specific choices of them -- regarded as fine-tuning -- are allowed by observational data.%
\footnote{See also a recent paper \cite{deRham:2018red} for an attentive discussion on the validity of effective theories.}
Let us elaborate further on the latter fine-tuned models in this section. Note that the fine-tuning is chosen for a given cosmological background (including all matter content). For the same reason, the fine-tuning is sensitive to inhomogeneities on smaller scales. The departure from $c_T^2=1$ is to be proportional to the relative over/under-density and higher $X$ derivatives of $G_4$ and $G_5$. In what follows, we speculate on possible constraints due to the presence of inhomogeneities even if the theory is fine-tuned to have $c_T^2 =1$ on the homogeneous background. We consider two cases, a self-accelerating generalized Proca (i.e. $\dot H =\dot\phi=0$) and the tracker solutions proposed in Ref.~\cite{DeFelice:2016yws}. The latter solutions are found for the following form of the generalized Proca functions:
\begin{align}
G_{2}(X)=b_2 X^{p_2}\quad&,\quad G_{3}(X)=b_3 X^{p_3}\,,\\ G_{4}(X)=\frac{M_{\rm Pl}^2}{2}+b_4 X^{p_4}\quad&{\rm and} \quad G_{5}(X)=b_5 X^{p_5}\,,
\end{align}
where
\begin{align}
 2p_3=p+2p_2-1\quad,\quad p_4=p+p_2\quad{\rm and}\quad 2p_5=3p+2p_2-1\,.
\end{align}
The relations among the powers of $X$ are inferred by inspection of Friedmann equations. For later use, let us introduce the following dimensionless variables
\begin{align}
y\equiv \frac{b_2\phi^{2p_2}}{3M_{\rm Pl}^2H^2 2^{p_2}}\quad {\rm and}\quad \beta_i\equiv \frac{p_ib_i}{2^{p_i-p_2}p_2b_2}\left(\phi^pH\right)^{i-2}\qquad (i=3,4,5)\,,
\end{align}
where $\phi$ is the homogeneous background of $A_\mu$ \eqref{vectorvev}. Using these variables, the Friedmann and field equations solve $y$ and $\beta_3$ in terms of $\beta_4$, $\beta_5$ and $\Omega_{m}\equiv\frac{\rho_{m}}{3H^2M_{\rm Pl}^2}$. Namely, we have that
\begin{align}\label{eq:beta2}
y=\frac{p_2\left(p+p_2\right)}{\beta}\left(1-\Omega_{m}\right) \quad{\rm and}\quad 1+3\beta_3+6\left(2p+2p_2-1\right)\beta_4-\left(3p+2p_2\right)\beta_5=0\,,
\end{align}
where $\beta\equiv -p_2\left(p+p_2\right)\left(1+4p_2\beta_5\right)+6p_2^2\left(2p+2p_2-1\right)\beta_4$.

\subsection{General expectation}

Naively, we expect that even if the model is fine-tuned to $c_T^2=1$ on cosmological backgrounds, such fine-tuning cannot account for inhomogeneities. This is easily understood if we look at the fine-tuning relations. For the self-accelerating model ($\dot{H} = \dot\phi = 0$) we have that $c_T^2=1$ requires, from \eqref{eq:tensors},
\begin{align}\label{eq:ct1}
G_{4,X}=\frac{1}{2}{H\phi} \, {G_{5,X}}\,.
\end{align}
In this case, if we approximate an inhomogeneity by a local shift of the background, say $H\to H +\delta H$ due to $\Omega_m\to \Omega_m+\delta \Omega_m$, it is clear that the relation \eqref{eq:ct1} will not hold, due to the change in $H$, or that the fine-tuning has to be extended to higher orders counteracting such a local shift, which would require different fine-tunings of the functions at different locations of the universe and thus be would unreasonable. This fact is even clearer if we look at the condition $c_T^2=1$ in the tracker solutions, that is
\begin{align}\label{eq:beta4}
\beta_4=\frac{\beta_5}{4}\frac{2p_1+2p_2+\Omega_{m}\left(3+2p_2\right)}{p_1+p_2-p_2\Omega_m}\,.
\end{align}
In this form, it is obvious that any inhomogeneity in $\Omega_m$ breaks the fine-tuning. Let us estimate the effects of such inhomogeneities and see that indeed fine-tuned theories are still subject to stringent constraints. Note that fine-tunings that are not background dependent, like a subset in DHOST theories \cite{Langlois:2017dyl}, are not subject to such constraint in the same fashion.

\subsection{Environmental $\delta c_T$ due to $\delta\Omega_{\rm m}$}
We can obtain a qualitative estimate of the shift in the GW propagation speed by first assuming that the condition $c_T^2=1$ holds and then by expanding around the solution obtained under this condition for a (small) change in the matter energy density. Such an expansion involves higher derivatives with respect to (w.r.t.) $X$ of $G_4$ and $G_5$, which we assume to be parametrically small with a parameter $\epsilon\ll 1$ (consistent with the tight bound from observations). With these assumptions, the change in $c_T^2$ can be estimated in general to be
\begin{align}\label{eq:deltact}
\delta c^2_T=-\epsilon \,C\,\delta\Omega_m\,,
\end{align}
where the constant $\epsilon\, C$ depends on the specifics of the underlying model. In particular, for the self-accelerating solution we have that
\begin{align}
\epsilon \, C_{\rm s.a.}\equiv\frac{1}{2g_{40}}\frac{g_{51}\left(3g_{32}+g_{22}\right)+g_{31}\left(6g_{42}-3g_{52}\right)}{3\left(g_{31}+4g_{40}\right)g_{31}+4g_{40}\left(g_{22}+3g_{32}\right)}\,,
\end{align}
where we have introduced a series of dimensionless parameters to quantify the magnitude of each derivative w.r.t.~$X$ as
\begin{align}
 g_{2n}\equiv G_{2,X^n} \frac{\phi^n}{M_{\rm Pl}^2H^2}\quad,\quad
 g_{3n} \equiv G_{3,X^n} \frac{\phi^n}{M_{\rm Pl}}\,,\\
 g_{4n} \equiv G_{4,X^n}  \frac{\phi^n}{M_{\rm Pl}^2}\quad ,\quad
 g_{5n} \equiv G_{5,X^n} \frac{H^2\phi^n}{M_{\rm Pl}}\,,
\end{align}
and we have used the background equations of motion as well as the $c_T^2=1$ condition. We have assumed that $g_{4n}$ and $g_{5n}$ with $n>0$ are small and of order $\mathcal{O}(\epsilon)$. On the other hand, for the tracker solutions we find
\begin{align}
\label{Ctracker}
\epsilon \, C_{\rm tracker}\equiv\beta_5\frac{9p_2\left(p+p_2\right)\left(1-\Omega_{m}\right)}{\left(p+p_2\left(1-\Omega_{m}\right)\right)^2}\,.
\end{align}
As we expected, $\epsilon \,C$ depends in general on first $X$ derivatives of $G_5$ and second $X$ derivatives of $G_4$ and $G_5$. Without assuming any further fine-tuning, it is reasonable to assume that $C\sim O(1)$ or not much less than that. Note that this is not unique to the generalized Proca theory and a similar expansion would hold for those scalar-tensor theories that modify the GW propagation speed as well. We can use the current bound on the GW propagation speed to constrain the value of $\epsilon$.

\subsection{Rough constraint}

The binary Neutron Star merger associated with GW170817/GRB170817A is believed to have happened at the galaxy NGC4993 \cite{Hjorth:2017yza,Im:2017scv} which is around $40 \, {\rm Mpc}$ away from us. This means that photons took roughly $4\times10^{15} \, {\rm s}$ to arrive on the earth after their emission. If we assume that GWs propagated with a constant speed during the whole path, the fact that the detection of GWs and that of $\gamma$-rays coincided within $\mathcal{O}(1) \, {\rm s}$ tells us that $\delta c_T\sim 10^{-15}$ in units of $c_{\gamma}=1$, as in \eqref{eqn:GW170817constraint}. For a fine-tuned model, we have $c_T = 1$ on the homogeneous cosmological background, and thus we have to account for the modification of $c_T$ only inside the lumps of dark matter, or the vector field if it clumps. This is obviously not easy to model but we can build a simple yet reasonable estimate and see that even in that case the constraint is not changed much from more accurate estimates. Let us consider a rough model where the propagation speed of GWs is only modified when they go through an over/under-density. As a simple exercise, we can compute what would be the effect if they only encountered a single object of over-density. In that case, the delay (or the advance if $\Delta t_{\rm delay}$ defined below is negative) only depends on how large the over-density is and the distance traveled by the GWs inside the object, say $L$, so that formally we can write
\begin{align}
\Delta t_{\rm delay}\equiv \Delta t_{\rm GW}-\Delta t_{\rm photon}=L\,\delta c_{T} 
\, < \, \mathcal{O}(1) \, {\rm s}
\end{align}
where $\delta c_T$ is given by Eq.~\eqref{eq:deltact}. In that case, the bound on the $\epsilon$ parameter is given by
\begin{align}
\epsilon\, C <{2\times10^{-14}}\Delta^{-1} \left(\frac{1 \rm Mpc}{L}\right)\,,
\end{align}
and we have introduced $\Delta\equiv{\delta\rho_{\rm halo}}/{\rho_c}$, where $\rho_c$ is the critical energy density of the universe. Assuming $C \sim O(1)$ (see the texts below \eqref{Ctracker}), the effects due to halos of different sizes can be found in Table \ref{tab:examples}. For illustrative purposes, we consider a typical super-cluster, a cluster of galaxies, a galaxy and a galaxy inside a cluster. However, the exact properties of dark matter halos are yet to be fully understood. For example, how one should define a halo is a subtle issue. Thus, average densities, radius and mass of the halos might depend on the definitions one uses. Here we use the spherical overdensity definition in which a ``virialised'' halo is defined as a sphere in which the average density is $\Delta=200$ times the critical density \cite{Lacey:1994su}. This is a useful definition based on spherical collapse and widely used in numerical simulations. A substructure has a higher value for the overdensity due to tidal disruptions \cite{Diemand:2007qr,GonzalezCasado:1994dx}. To compute the average density of a substructure (a galaxy inside a cluster) we assume that the profile of the subhalo is truncated where its density is equal to the local density inside the cluster. Thus, the closer to the center of the main halo the larger the $\Delta$ for the subhalo. To estimate that, we use the Navarro-Frenk-White profile \cite{Navarro:1995iw} and we use the values from Ref.~\cite{Duffy:2008pz}. We find that typically inhomogeneities lead to a constraint $\epsilon<10^{-16}$. Note that here we do not take into account that the halo has an extended profile and, thus, we are overestimating the effect.
 \begin{table}
  \begin{center}
   \begin{tabular}{||c |c c c c||}
    \hline
    &
    $M_{\Delta}(h^{-1}M_\odot)$ & $r_{\Delta}(h^{-1}{\rm Mpc})$ & $\Delta$ & $\epsilon$ \\ [0.5ex] 
    \hline\hline
    super-cluster (SC) &
	$10^{15}$ & $1.6$ & $200$ & $4\times 10^{-17}$ \\ [0.5ex] 
    \hline
    cluster &
	$10^{14}$ & $0.7$ & $200$ & $10^{-16}$ \\
    \hline
    galaxy &
	$10^{12}$ & $0.16$ & $200$ & $4\times 10^{-16}$ \\[0.5ex] 
    \hline
    galaxy in SC &
	$5\times 10^{11}$ & $0.08$ & $1000$ & $10^{-16}$ \\
    \hline
   \end{tabular}
   \caption{Modification of the propagation speed of GWs due to non-linear structure and the corresponding constraint on $\epsilon$ from GW170817/GRB170817A. From top to bottom we consider a super-cluster, a cluster, a galaxy and a galaxy inside a supercluster. We use the spherical overdensity $\Delta=200$ to define the virial radius and mass. Substructure (e.g.~the halo of a galaxy inside a cluster) has a larger spherical overdensity because of tidal disruptions, as they are sitting on top of the main halo. It may range from $\Delta_{SH}=300 - 1000$ depending on whether the substructure is closer to the edge or the center of the main halo. We find approximately $\Delta_{SH}=300$ and $\Delta_{SH}=1000$ for a galaxy inside a cluster of $M_\Delta=10^{15}h^{-1}M_\odot$, respectively located at $0.8\,r_\Delta$ and $0.5\,r_\Delta$.}
   \label{tab:examples}
  \end{center}		
 \end{table}

We can slightly refine the model by assuming that the dark matter lumps are uniformly randomly distributed along the line of sight, $D=40 \, {\rm Mpc}$, since the real distribution is rather unknown.
Let us divide $D$ into boxes of size $L$, each box containing a lump of size $R$ with a given $\delta c^{2}_T\propto -\epsilon \delta\Omega_m$, which can be either positive (under-density) or negative (over-density) with equal probability. We have $D^3/L^3$ total objects with a cross section of $\sigma=4\pi R^2/L^2$ and $N=D/L$ objects in the line of sight. Therefore, we effectively have $n_E=N \sigma$ encounters.
The time delay due to an encounter is given by
\begin{align}
\Delta t_{\rm delay,E}\equiv \Delta t_{\rm GWs,E}-\Delta t_{\rm photon}=R\,\delta c_{T}
\end{align}
Due to the assumption that they are uniformly distributed we have $\langle\Delta t_{\rm delay}\rangle=\sum_i\Delta t_{\rm delay,i}=0$ (they average to the mean background density for which fine-tuning takes place). Nevertheless, we know that for random walks the variance is non-zero, i.e.
\begin{align}
\sqrt{\langle\Delta t_{\rm delay}^2\rangle}=R\,\delta c_{T} \sqrt{n_E}\,.
\end{align}
Let us consider a favorable case. For example, if the average lump that the GWs find has $R=100 \, {\rm kpc}$ (galaxy size) separated by $L=1 \, {\rm Mpc}$ (typical separation) we get $n_E\sim 5$, which yields a constraint $\epsilon< 2\times 10^{-16}$. Since we expect these objects not to be very dense, we conclude that the constraint presented here does not change its order of magnitude even if we take into account multiple galaxies/clusters along the line of sight. It should be noted that the rough calculation presented here is over-estimating the effects of the inhomogeneity since we have assumed that $C\sim 1$ and that the average density of the halo is given by the spherical overdensity $\Delta=200$, estimated from the spherical collapse. We emphasize however that the main point is to show that even if one fine-tuned a theory to have $c_T^2=1$ on the cosmological background it would likely be ruled out due to non-linear effects of inhomogeneous structures in general unless these effects are somehow suppressed as well. Actually, the GW at least propagated through a part of NGC4993 and a part of our galaxy and thus, according to Table~\ref{tab:examples}, this already puts a stringent bound on $\epsilon$.

 \section{Summary and discussions}
\label{sec:summary}

In this work we studied the effect of a disformal transformation with constant factors on the generalized Proca theory and the implications concerning the sound speeds of the perturbations on cosmological backgrounds. Such consideration is timely, since the simultaneous detections of gravitational waves and gamma rays, credibly originated from the same neutron star merger, have placed a stringent constraint \eqref{eqn:GW170817constraint} on the propagation speed of GW in space. This in turn highly restricts the range of modifications of the gravity theory that deviates the GW propagation speed from that of light. Since measurements of gravitational waves are affected not only by a given (modified) gravity theory but also by how matter interacts with them, it is of natural interest to ask how a metric transformation affects the propagation speed of the gravitational degrees of freedom. The generalized Proca theory involves a massive vector field in the gravity sector in addition to the standard spin-$2$ gravitons. Metric transformations of this theory can thus exhibit a highly non-trivial structure, and the understanding of it was the main purpose of our study.

In Section~\ref{sec:Proca}, we introduced the action \eqref{eq:genProc} of the generalized Proca theory \cite{Heisenberg:2014rta,Allys:2015sht,Jimenez:2016isa,Allys:2016jaq}, as well as disformal transformation \eqref{disformal} of the metric. By explicitly computing the transformation laws of the full Lagrangian, we found that it is closed under such transformations with constant conformal $\Omega$ and disformal $B$ factors, namely the action after transforming the generalized Proca action \eqref{eq:genProc} reduces to the same form only with redefinitions of the functions in $\mathcal{L}_{2,3,4,5,6}$ (\ref{ProcL2}--\ref{ProcL6}). In this sense, disformal transformations with constant factors are the ones that are compatible, hence naturally associated, with the generalized Proca theory. The redefinition of the functions in $\mathcal{L}_{2,3,4,5,6}$ mix them with each other, as illustrated in FIG.~\ref{fig:diagram}. This is in contrast to the Horndeski theory where each Lagrangian is closed under a constant disformal transformation. The main difference is the existence of antisymmetric terms proportional to $F_{\mu\nu}$ arising in the transformation, which are absent in the case of the Horndeski theory and can be seen by the substitution $A_\mu = \nabla_\mu \phi$. In fact, the closure under such transformations holds only provided that we include an extra term proportional to $\tilde{F}^{\mu\rho} \tilde{F}_{\nu\sigma} A_\mu A^\nu \nabla_\rho A^\sigma$ in \eqref{ProcL5}, see also (\ref{eq:newG5}), which was until now only present in the so-called ``beyond generalized Proca'' theories \cite{Heisenberg:2016eld}. Also, the function $G_2$ in ${\cal L}_2$ \eqref{ProcL2} needs to depend on $U^2$, where $U = - F_{\mu\nu} \tilde{F}^{\mu\nu} /4$, and $Y = F_{\mu\rho} F_\nu{}^\rho A^\mu A^\nu$, as well as $X = - A_\mu A^\mu / 2$ and $F = - F_{\mu\nu} F^{\mu\nu} / 4$. Looking at the transformation of each part of the action, $S_{2,3,4,5,6}$, we observed that $S_3$ and $S_5$ are individually closed under the considered transformation (given the aforementioned term $\tilde{F} \tilde{F} A A \nabla A$ is present), while $S_4$ and $S_6$ produce extra terms contributing to $S_2$ after the transformation. This behavior is in part naturally expected due to the fact that $S_{2,4,6}$ are even under the change $A_\mu \to - A_\mu$ and $S_{3,5}$ are odd while the disformal transformation preserves this nature, and by counting the number of derivatives in each term. We verified this explicitly and obtained the actual forms of the changes of the functions and of those extra terms. For convenience, we summarized the transformation rules in Section~\ref{subsec:trans_summary}.

In Section~\ref{sec:soundspeeds}, the sound speeds of the scalar, vector and tensor modes of the perturbations around the flat FLRW background were discussed. We first provided their expressions, which are mostly in the existing literature \cite{DeFelice:2016yws} except the newly added term $\propto {\cal G}_5$, and then showed how basic quantities transform, before calculating the resulting transformation rules of the sound speeds. Our main finding is that all the scalar, vector and tensor sound speeds transform identically as $c_s^2 \to c_s^2 / (1 - 2 B X)$. This can in fact be understood as the change in the lightcone structure. While a conformal transformation -- the factor $\Omega$ in \eqref{disformal} -- does not affect the lightcone, the disformal part $B$, on the other hand, widens/narrows the lightcone and changes $c_s^2$ between the two frames. For this reason, the changes in the speed of propagation coincide among all the modes.
Let us note however that the metric transformation is a change of variables (as long as it is invertible) for a given theory, and thus once the action of the whole system including the matter sector and its coupling to gravity is specified, the difference in $c_s^2$ between the two frames are not physically measurable quantities. What is observable is e.g.~the ratio $c_{S,V,T}^2/c_{\gamma}^2$, which is actually invariant under the disformal transformation. On the other hand, if the matter sector is minimally coupled to the metric after the transformation then the ratio $c_{S,V,T}^2/c_{\gamma}^2$ is different from that for the case where the matter is coupled to the metric before the transformation. Alternatively, the difference in $c_s^2$ between the two frames becomes observable if, e.g., one matter component couples to one metric $g_{\mu\nu}$ while another to the other $\bar{g}_{\mu\nu}$ and if one can retrieve the information from both of them.

Once we understood the structure of generalized Proca theory that is closed under a class of metric transformations, we discussed the consequences of the observational constraint on the propagation speed of gravitational waves. The tight constraint on the propagation speed tells us that either there is no derivative coupling to gravity (i.e. the terms $G_4$ and $G_5$) or that there is a severe fine-tuning between the functions $G_4$ and $G_5$. Such a tuning has to be made for a given background as the speed of gravitational waves depends on $\dot\phi$ and $H$. We took the fine-tuning carefully and argued, with rough estimates, that any inhomogeneity would drive the solution out of the fine-tuning simply because locally an inhomogeneity can be thought as a shift in the background. In this way, the departure from $c_T=1$ is proportional to the over/under-density of the inhomogeneity and derivatives of $G_4$ and $G_5$ w.r.t.~$X$. The deviation from $c_T=1$ would only occur inside the inhomogeneities and, thus, we recast the constraint of $c_T=1$ for such a case. We considered a toy model where the inhomogeneities are dark matter halos with spherical overdensity $\Delta=200$, uniformly distributed such that they average out to the mean background density. However, like in a random walk, the standard deviation is non-vanishing and we expect to see a departure from $c_T=1$. We have seen that, for the tuning of $c_T=1$ to hold, one needs to further fine-tune the functions $G_4$ and $G_5$ so that they cancel at higher orders as well or that they are as small as $10^{-16}$. Although our estimate is based on some crude assumptions, the fact that the constraint is still very tight led us to expect that similar results would hold even for a more accurate estimate. We concluded then that the fine-tuning has to take place independently of the background, like a subclass of DHOST theories \cite{Langlois:2017dyl}, otherwise some further tuning is required.

Throughout the present paper we have studied classical properties of the generalized Proca Lagrangian and the speed of propagation of gravitational waves by analyzing their transformation rules under disformal transformations. It is certainly worthwhile investigating their quantum properties. Especially, it is important to ask in which class of theories a small $\delta c_{\rm T}$ is technically natural by computing loop corrections to $\delta c_{\rm T}$. Another intriguing possibility is to invoke a (partial) UV completion that recovers $\delta c_{\rm T}\simeq 1$ at the LIGO frequencies, taking advantage of the low cutoff scale of an effective field theory~\cite{deRham:2018red}.

\section*{Acknowledgements}
G.D. would like to thank E. Salvador-Sol{\'e} for useful discussions on dark matter halos and their substructure, and R.N. is grateful to Daisuke Yoshida for discussions on transformations in generalized Proca theories. The work of S.M. was supported by Japan Society for the Promotion of Science (JSPS) Grants-in-Aid for Scientific Research (KAKENHI) No. 17H02890, No. 17H06359, and by World Premier International Research Center Initiative (WPI), MEXT, Japan. 
R.N. was supported by the Natural Sciences and Engineering Research Council (NSERC) of Canada and by the Lorne Trottier Chair in Astrophysics and Cosmology at McGill University. G.D. acknowledges the support from DFG Collaborative Research centre SFB 1225 (ISOQUANT). G.D. and R.N. thank the Yukawa Institute for Theoretical Physics (YITP) at Kyoto University for its hospitality during the progress of this work. Discussions during the workshop YITP-T-17-02 on ``Gravity and Cosmology 2018'' and the YKIS2018a symposium on ``General Relativity -- The Next Generation --'' were particularly useful to complete this work. V.P. would also like to thank the YITP for welcoming him for an internship where this work was carried out. Involved calculations were cross-checked with the \texttt{Mathematica} {package} \texttt{xAct} (www.xact.es).

\bibliography{vectordisformal.bib}

\end{document}